# *Euclid* Mission: building of a Reference Survey

J. Amiaux[a], R. Scaramella[b], Y. Mellier[c], B. Altieri[d], C. Burigana[e], A. Da Silva[f], P. Gomez[d], J. Hoar[d], R. Laureijs[g], E. Maiorano[e], D. Magalhães Oliveira[f], F. Renk[h], G. Saavedra Criado[g], I. Tereno[i], J.L. Auguères[a], J. Brinchmann[j], M. Cropper[k], L. Duvet[g], A. Ealet[l], P. Franzetti[m], B. Garilli[m], P. Gondoin[g], L. Guzzo[n], H. Hoekstra[j], R. Holmes[o], K. Jahnke[o], T. Kitching[p], M. Meneghetti[q], W. Percival[r], S. Warren[s], and the *Euclid* collaboration

[a]Commissariat à l'Energy Atomique, Orme des Merisiers, 91191 Gif sur Yvette, France; [b]INAF Osservatorio Astronomico di Roma, Via Frascati 33, 00040 Monteporzio Catone (Roma), Italy; [c]Institut d'Astrophysique de Paris, 98 bis Boulevard Arago, 75014 Paris, France; [d]European Space Agency, ESAC, POB 78, 28691 Villanueva de la Cañada, Madrid, Spain; [e]INAF-IASF Bologna, I-40129 Bologna, Italy; [f]Centro de Astrofísica, Universidade do Porto, Rua das Estrelas, 4150-762 Porto, Portugal; [g]European Space Agency, ESTEC, Keplerlaan 1, 2200AV Noordwijk, The Netherlands; [h]European Space Agency, ESOC, Robert-Bosch-Str. 5, 64293 Darmstadt Germany; [i]Centro de Astronomia e Astrofisica da Universidade de Lisboa, Tapada da Ajuda, 1349-018 Lisboa, Portugal; [j]Leiden Observatory, Leiden University, Niels Bohrweg 2, 2333, The Netherlands; [k]Mullard Space Science Laboratory, University College London, Holmbury St Mary, Dorking Surrey RH5 6NT, United Kingdom; [l]Centre de Physique des Particules de Marseille, 163 Avenue de Luminy, Case 902, 13288 Marseille Cedex, France; [m]INAF IASF, Milano Via E Bassini 15, 20133, Italy; [n]INAF- Osservatorio Astronomico di Brera, Via Bianchi 46, 23807 Merate, Italy; [o]Max-Planck-Institut für Astronomie, Königstuhl 17, D69117 Heidelberg, Germany; [p]Royal Observatory Edinburgh, Blackford Hill, EH9 3HJ Edinburgh, United kingdom; [q]INAF - Osservatorio Astronomico di Bologna, Via Ranzani 1, I-40127 Bologna, Italy; [r]University of Portsmouth, Burnaby Road, PO1 3FX Portsmouth, United Kingdom; [s]Imperial College London, Blackett Laboratory, Prince Consort Road, London SW7 2AZ, United Kingdom

## ABSTRACT

*Euclid* is an ESA Cosmic-Vision wide-field-space mission which is designed to explain the origin of the acceleration of Universe expansion. The mission will investigate at the same time two primary cosmological probes: Weak gravitational Lensing (WL) and Galaxy Clustering (in particular Baryon Acoustic Oscillations, BAO). The extreme precision requested on primary science objectives can only be achieved by observing a large number of galaxies distributed over the whole sky in order to probe the distribution of dark matter and galaxies at all scales. The extreme accuracy needed requires observation from space to limit all observational biases in the measurements. The definition of the *Euclid* survey, aiming at detecting billions of galaxies over 15 000 square degrees of the extragalactic sky, is a key parameter of the mission. It drives its scientific potential, its duration and the mass of the spacecraft. The construction of a Reference Survey derives from the high level science requirements for a Wide and a Deep survey. The definition of a main sequence of observations and the associated calibrations were indeed a major achievement of the Definition Phase. Implementation of this sequence demonstrated the feasibility of covering the requested area in less than 6 years while taking into account the overheads of space segment observing and maneuvering sequence. This reference mission will be used for sizing the spacecraft consumables needed for primary science. It will also set the framework for optimizing the time on the sky to fulfill the primary science and maximize the *Euclid* legacy.

# 1. INTRODUCTION

*Euclid* is the M2 ESA cosmic vision mission, selected in October 2011 and adopted in June 2012. The primary goal of this mission is to map the geometry and evolution of the dark universe with unprecedented accuracy and precision in order to place stringent constraints on Dark Energy, Dark Matter, Gravity and cosmic initial conditions[1]. This mission will use two independent cosmological probes: Weak gravitational Lensing (WL) and Galaxy Clustering (GC).

The following topics with their key questions are addressed by *Euclid*:
1. Dynamical Dark Energy: Is the dark energy simply a cosmological constant, or is it a field that evolves dynamically with the expansion of the Universe?
2. Modification of Gravity: Alternatively, is the apparent acceleration instead a manifestation of a breakdown of General Relativity on the largest scales, or a failure of the cosmological assumptions of homogeneity and isotropy?
3. Dark Matter: What is dark matter? What is the absolute neutrino mass scale and what is the number of relativistic species in the Universe?
4. Initial Conditions: What is the power spectrum of primordial density fluctuations, which seeded large-scale structure, and are they described by a Gaussian probability distribution?

In order to address those questions, *Euclid* will measure the shape and spectra of galaxies over the best 15 000 deg² of the extragalactic sky in the visible and Near Infrared, out to redshift about $z = 2$, thus covering the period over which dark energy took over and started accelerating the universe expansion.

The payload baseline comprises two wide field instruments (>0.5 deg²): a visible instrument (VIS), and a near infrared photometric and spectroscopic instrument (NISP P and NISP S). The visible channel is used to measure the shapes of galaxies for weak lensing, with a platescale of 0.100 arcsec in a wide visible red band (R+I+Z, 0.55 to 0.92 µm). The NIR photometric channel provides three NIR bands (Y, J, H, spanning 1.1 to 2.0 µm) with a platescale of 0.300 arcsec.
The baseline for the NIR spectroscopic channel operates in the wavelength range 1.0 to 2.0 µm in slitless mode at a spectral resolution *R*=250 for a 1 arcsec diameter source, with 0.300 arcsec per pixels.

This paper describes the way the observations with these instruments are collected during the survey in order to achieve the major science goals, taking into account the constraints to the mission.

# 2. *EUCLID* SURVEY SCIENCE DRIVERS

In order to meet *Euclid* top level Science Case, the Wide Survey must ensure that the following parameters are properly sampled by both the imaging and spectroscopic probes:
- density of objects for which required Signal to Noise Ratio (*SNR*) is reached
- volume probed for cosmic variance

The volume, for fixed depth, translates into a survey area. The pattern on the sky covered by *Euclid* observations must sample all relevant angular scales that probe signatures of dark energy in the correlation functions and angular power spectra (scales between $l\sim$ some 10 and $l\sim$5000). Requirement Requirement for depth to probe redshift from $z = 0.7$ to $z = 2.0$ translates into requirements on the WL galaxies observation in the visible to $m_{AB}$= 24.5 at 10-σ, and in the near infrared to $m_{AB}$= 24.0 at 5-σ, and on the galaxy spectra to *SNR* for $H_\alpha$> 3.5 for $H_\alpha$ flux $\geq 3\ 10^{-16}$ erg cm$^{-2}$s$^{-1}$, see paper R. Laureijs et al. in these proceedings (2012)[2].

The optimum functioning point of the *Euclid* Mission has been set to the following parameters:
- a joint 15 000 deg² survey of the extragalactic sky
- an average 30 of galaxies/arcmin² from imaging channels useful for Weak Lensing probe
- an average 3500 of galaxies/deg² from spectroscopy channel useful for Galaxy Clustering probe

End-to-end simulations have been performed in order to define which part of the extragalactic sky would be most useful for both probes in order to meet the three driver requirements. The contributors of main importance to galaxy counts - zodiacal background and extinction affecting the *SNR*, bright stars affecting the spectra contamination and star saturation - have been used as free parameters in the end-to-end simulations. The survey parameters related to spacecraft, including instruments parameters were set to the reference value of the definition phase (these values include the required margin set by ESA ECSS standard at definition phase).

The results of these analyses are setting the limits of the extragalactic survey and help defining the strategy. The survey will be built starting observations on the core preferred area starting from the ecliptic poles, where the zodiacal background is minimum[3]. Whenever survey coverage efficiency drops on the core area the survey operation will cover the extended preferred area. Best coverage of those areas will ensure maximum efficiency of the survey and fulfillment of survey driving requirements on scales and galaxy number density. Core and Extended preferred area limits are defined as in table 1:

|  | Core preferred area | Extended preferred area |
| --- | --- | --- |
| Zodiacal light: Ecliptic latitude | >|15°| | >|15°| |
| Star Density: Galactic latitude | >|25°| | >|20°| |
| Extinction: E(B-V) | 0.08 | 0.15 |
| Total Area | 14 700 deg² | 18 400 deg² |

Table 1. *Euclid* definition of Core preferred area and Extended preferred area. The survey starts from the fields that have the highest object density and move towards the geometrical limits set by Zodiacal light (Ecliptic cut), Extinction (Contour) and star density (Galactic latitude).

## 3. REFERENCE SEQUENCE OF OBSERVATION

The survey is built in step-and-stare mode, meaning that the spacecraft will point toward a designated field on the sky, perform a nominal sequence of observations that includes visible imaging mode, near infrared photometry imaging mode and near infrared spectroscopic mode. The details of these modes are described in the associated papers by M. Cropper et al.[4] and E. Prieto et al.[5] in these proceedings (2012). Each field is observed with a sequence of 4 frames, the spacecraft performing small dithering steps between each frame (~100 arcsec). The sequence of dithering is designed in order to recover the gaps between the detectors on the visible and near infrared instrument focal planes. The target signal to noise ratio is obtained after 3 frames of observation, and the dithering strategy ensures that 95% of the pixels in the visible and more than 90% of the pixels in the near infrared will be observed with at least 3 frames, 50% of all pixels will be observed with more than 3 frames.

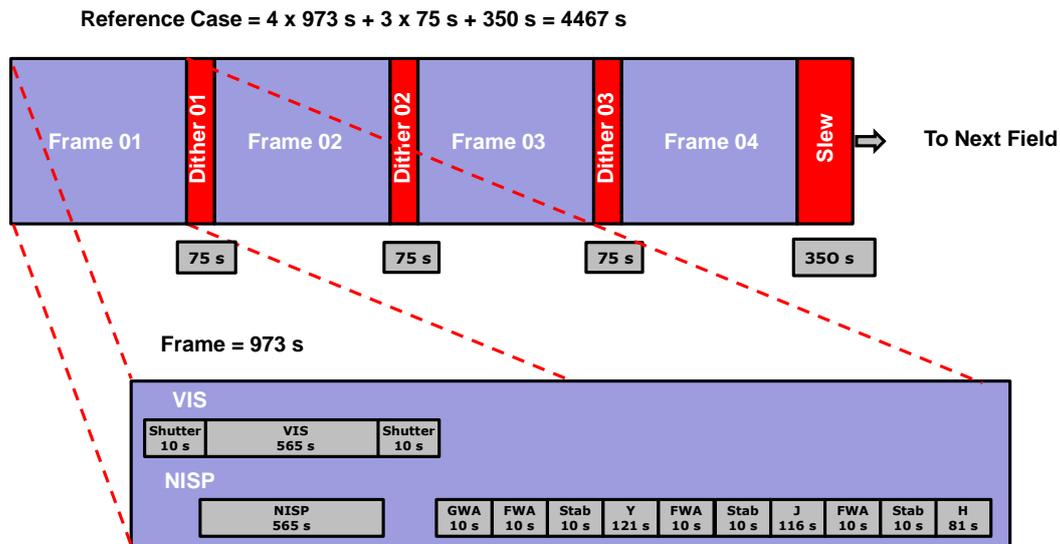

Figure 1. *Euclid* reference science field sequence of observation.

During the construction of the Wide survey, the reference operation sequence must ensure that the all calibration fields are observed. The final capability to remove systematics effect will be set by the way we control them by design on the spacecraft, and correct for the residual by extensive and appropriate calibrations of data.

The sequence must also ensure the construction of the *Euclid* Deep Field, which will be revisited over the years, primarily for calibration purpose. It will be completed by additional revisits in order to ensure that the Deep field is observed 2 magnitudes deeper than the Wide Survey. The Deep survey will have a size of 40 deg² (which can be segmented) and will require a total of 40 visits with the nominal *Euclid* observation sequence.

The calibration sequence is set by the calibration needs of visible imaging, near infrared photometry, and near infrared spectroscopy, in order to achieve the higher level science requirements for Weak Lensing Tomography and Galaxy Clustering.

One of the main goals of the reference survey study was to verify whether the important calibration operations could be accommodated within the survey constraints. For all calibration operation sequences, we have allocated a specific calibration field locations and observation sequences that will be representative for survey duration and space segment consumable sizing. The exact fields and observation strategies will be refined in the later stages of the operation definitions.

For the visible channel, one of the main drivers is the knowledge of the full system PSF. The instrument is basically self-calibrated on science data, but will benefit from extra monthly PSF calibration on dense stellar fields, where the PSF can be intensively probed, including the monitoring of the degradation of detector charge transfer efficiency due to radiations. For the reference survey, the PSF calibration fields have been located in the galactic plane. An additional critical calibration for the visible instrument impacting the survey is the PSF color gradient effect. This calibration will be performed using a sample of HST high resolution images of galaxies. The same sample will be observed with *Euclid* in order to calibrate *Euclid* color gradient bias. This requires *Euclid* observations of some HST fields (which definition will be refined when the exact number of galaxies needed for calibration is precisely known). The current reference survey allows revisiting 3 times 3 different HST fields (COSMOS, GEMS and STAGES) and spend 1 day each time for calibration. This allocation includes at least a factor 2 margin relative to what is expected.

The driving requirement for the near infrared photometry channel calibration is the relative photometry accuracy of < 1.5% over the whole survey. A specific "cross calibration" strategy has been established that will ensure a global solution on relative photometry to be established using multiple observations of the same objects, or object with known flux on different parts of the detector, at different epochs of the survey. This approach has been derived from ground-based survey calibrations (see Padmanabhan et al [6]). This calibration sequence requires revisiting each month a few deg² field (< 5 deg²) with a specific slew pattern in order to maximize the fraction of the NISP focal plane over which the same objects are repeatedly observed. The sequence involves a few hours of near infrared photometry observations. The need to have a monthly revisit on the same field sets the location of this field to one of the Ecliptic poles that are the only places accessible all year round. In the current implementation, this field is located in the Northern Ecliptic Pole.

The near infrared spectroscopic channel will require specific calibrations, especially wavelength and spectro-photometric calibrations, to ensure that the required redshift accuracy is reached. First, a wavelength zero point determination is done using the direct images from the photometric channel , then the dispersion solution is computed by scanning a Planetary Nebulae or an open cluster on a grid of positions over all detectors in the fields. For the current implementation of Survey, Planetary Nebulae and Open Cluster in the galactic Plane will be observed at least twice a year. The relative spectrophotometric calibration will be maintained over the mission life time by observing HST standard fields during the Performance Verification phase and then regular basis. More secondary standards can be also defined in the Deep Field after a calibration transfer during the Performance Verification phase.

The last driving requirement in clustering is the need of a sample of observed galaxies (~140 000 objects) for which the spectra cross contamination can be calibrated to reach a purity criterion of 99% (this criterion sets the ratio between the number of galaxies that have a Hα line with sufficient SNR to be observed, and the number of galaxies for which a correct redshift measurement is achieved). It then imposes to have a dedicated field covering at least 40 deg² (that could be partitioned) observed at least 12 times with the reference, each sequence having a dispersion orientation varying by more than 10° to allow a good calibration of spectra contamination. The need to observe the same fields over the years with different observations enforces the selection of field locations close to the Ecliptic Poles. For the current implementation, the Calibration sample will be split in 2 x 20 deg² fields located close to the Northern and Southern Ecliptic Poles avoiding the Large Magellanic Cloud on the South and ensuring that a field is always accessible all year round.

The fact that 12 observations are performed on 2 x 20 deg² field for constructing a high purity spectroscopic sample clearly imposes to complete the Deep Survey on the locations of the 2 purity calibration fields in order to minimize the impact of the Deep Survey on the total Mission duration. Therefore, during the 6 years of the survey, 28 additional visits to those fields are planned in the reference survey in order to reach the 2 magnitude additional depth with respect to the Wide Survey.

Figure 2 below shows the typical location of targets assumed for the Reference survey implementation.

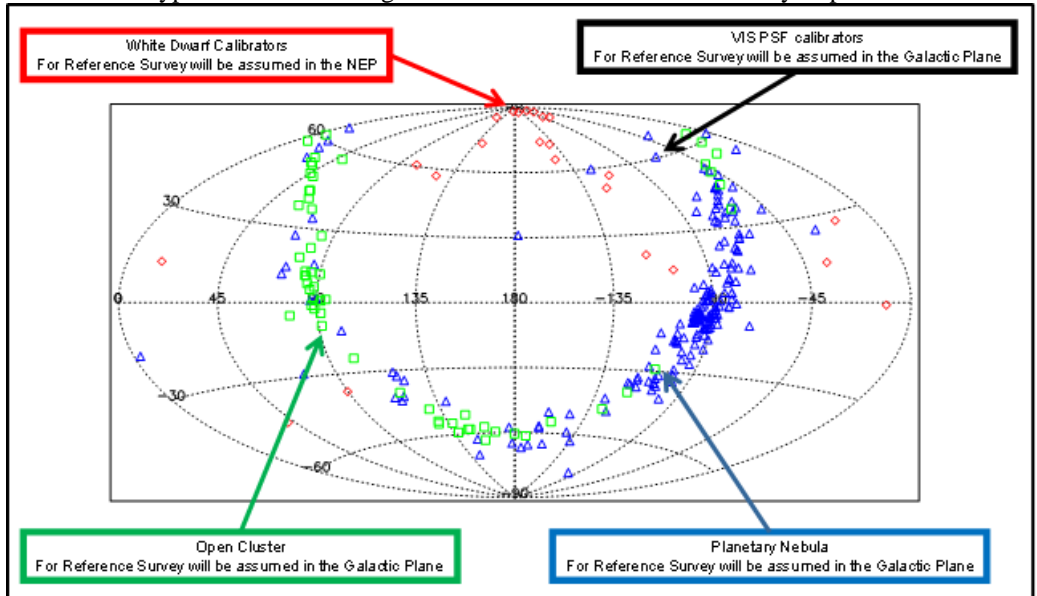

Figure 2. *Euclid* reference calibration objects location.

The final calibration objects selection will be done during planning of operation, but the density and location of objects in figure 2 show that Performance Verification phase will not depend on the launch date for availability of needed targets.

## 4. SPACECRAFT OPERATION KEY PARAMETERS

One main goal of the reference survey is to provide a sizing operation scenario, in order to have a best estimate of the needed capabilities at space segment level. These capabilities will be converted into spacecraft design implementation and must match the higher level constraints of the mission in terms of duration, mass allocation and Technology Readiness Level.

The main mission constraints are:
- **Mission duration**: nominal lifetime of 6.25 years including 6 years of nominal science operations.
- **Total Mass**: compatible with a launch on a single SOYUZ ST 2-1B from the Guiana Space Center
- **TRL**: Technology that have TRL ≥ 5 by the end of the Phase B1
- **Mission Operation**: The Ground Station Network shall be based on Cebreros and Malargue ESA ground stations.

The total mass allocated is a key parameter as it drives mainly the depointing capabilities (through the mass of the sunshield) and the number of field operations (if sizing of the Service Module is based on the assumption that the pointing actuation is performed using cold gas).

In order to ensure that the stability of the Payload Module is sufficient to ensure that the weak lensing measurement accuracy is not degraded by thermo-mechanical changes of the Payload Module, the spacecraft will be operated with its sunshield normal to the sun direction. Small variations around this attitude will be authorized within the following limitations:
- Maximum depointing range variation between any fields of the wide survey sequence shall be $\beta < 10°$
- Maximum absolute depointing shall be between 90° and 100° (the spacecraft is not allowed to point toward the sun for science observation to limit stray light).
- Maximum roll of the spacecraft shall be $-5° < \alpha < +5°$
- Maximum Solar Panel normal variation between 2 consecutive observation shall be $< 10°$
- The spaceraft shall be operable with depointing between 89° and 121°

The current sizing of the consumable for the pointing operations is based on cold gas approach. The Wide and Deep *Euclid* survey shall be performed in less than 42 000 science field operations (described in figure 1), and an equivalent of 36 large slews (< 180°). The spacecraft will be sized to be able to perform an additional 7$^{th}$ year of operations at the same operation rate as the wide survey.

The survey duration will need to fit into the duration allocation, taking into account the sequence of observations described in figure 1 of section 3. This sequence is composed of sub-exposures for acquisition of data in the different instrument channels. The remaining overheads are linked to space segment operations:
- Dither from 1 frame to the next (~100 arcsec): 75 s
- Slew from 1 field to the next (~0.75 deg): 350 s
- Slew from one patch to another (max 180°): <600 s
- Orbit station keeping maneuvers: 1 day / month (this is taken as a global bugdet of 2.5 additional months in the survey)
- Antenna pointing: no overheads

The definition of the mission Ground Station Network sets a limit to the maximum telemetry budget allocation that can be downloaded from the *Euclid* platform to the ground per day. Currently, the system shall be compatible with a K-band telemetry rate of 850 Gbits/day after payload science data compression. Taking into account the definition of the sequence of operations, the maximum number of sequences per day will be 19.2, leading to a maximum telemetry of:
- 180 Gbits/day for the NISP instrument
- 520 Gbits/day for the VIS instrument
- 700 Gbits/day for both instruments after data compression, which is compatible with the allocation.

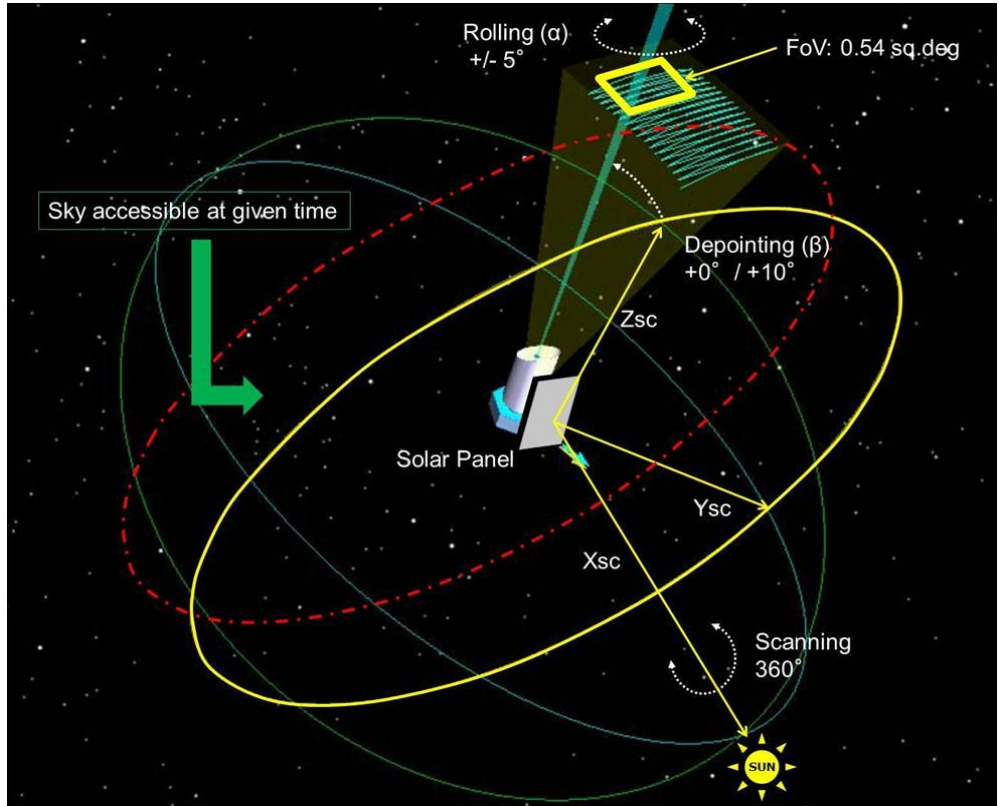

Figure 3. *Euclid* spacecraft pointing capabilities (Xsc, Ysc, Zsc gives orientation of S/C before any rotation).

All *Euclid* surveys will have to be verified with respect to the above parameters in order to ensure that the proposed scenarii are compatible with the space segment design.

## 5. SURVEY IMPLEMENTATION AND PERFORMANCE

The reference survey is built to demonstrate the feasibility of reaching the top level scientific requirements from galaxy measurements, over a broad range of scales based on obervations of the extragalactic sky, while the above described main constraints:

- **External** physical constraints on zodiacal light, extinction and star density
- **Calibration** constraint of revisiting fields with given cadence
- **Spacecraft** constraints limiting the pointing capability of the instruments, the number of operations and the survey duration

Table 2 below summarizes the main constraints that need to be taken into account while building the reference survey scenario.

| Constraint type | Constraint | Description |
|---|---|---|
| External | Zodiacal Light | Ecliptic lat >|15°| |
| External | Extinction | Contour E(b-v) < 0.08 –core |
|  |  | Contour E(b-v) < 0.15 –extended |
| External | Star density | Galactic lat >|25°| -core |
|  |  | Galactic lat >|20°| -extended |
| Calibration | VIS PSF calibration | Galactic Plane 1 day/month |
| Calibration | VIS PSF color gradient | HST fields COSMOS / GEMS / STAGES |
| Calibration | NISP P relative photometry | Revisit same 5sq.deg field 6 hours / month |
| Calibration | NISP S Wavelength | Planetary Nebulae / Open Cluster 1 day/ 6 months |
| Calibration | NISP S relative photometry | White Dwarf every 1 day / 6 months |
| Spacecraft | Survey Duration | <6 years (including 6 months margin) |
| Spacecraft | Field observation duration | ~4500 s |
| Spacecraft | Boresight depointing | +0° to +10° |
| Spacecraft | S/C Roll | -5° to +5° |
| Spacecraft | Panel Solar Aspect variation | <10° |
| Spacecraft | Number of field slew (primary) | <42 000 |
| Spacecraft | Orbital Maneuvers | 1 day/months |

Table 2. *Euclid* reference survey main constraints

The Survey is built using an ESAC developed tool, the *Euclid* Sky Survey Planning Tool (ESSPT) based on re-use of the Herschel operation planning tool.

The following sequence of images on figure 4 shows the building of the survey from year 1 to year 6 in cylindrical and orthographic projection. The sequence is built by covering first the regions with highest scientific value.

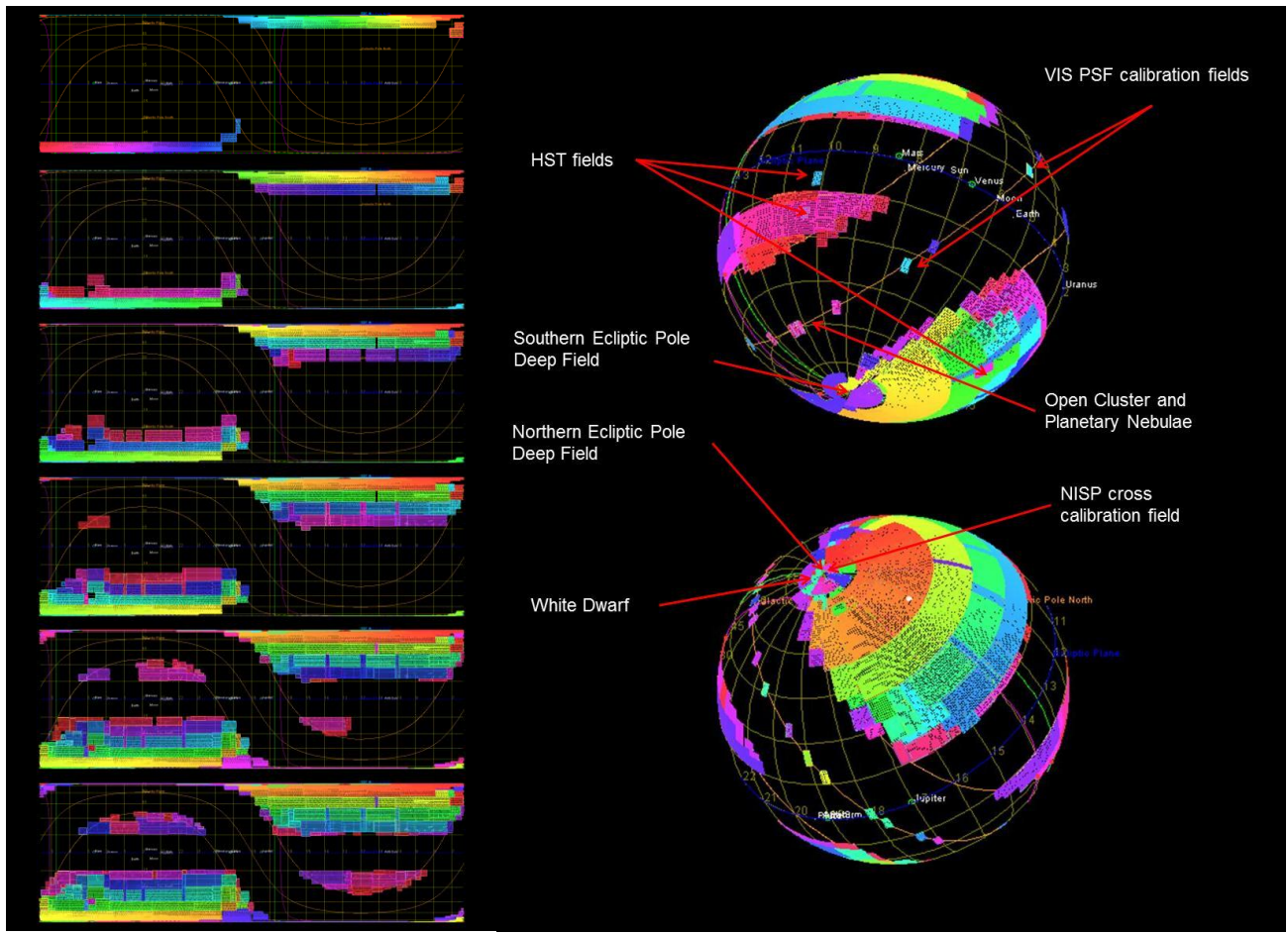

Figure 4. *Euclid* Reference Survey construction from year 1 (top) to year 6 (bottom) in cylindrical (left) and orthographic projection. In the left panel, only the wide survey fields are shown. In the right panel, the calibration fields are also displayed.

The ESSPT tool allows computing at each time of the sequence the S/C attitude and orbit parameters on a large amplitude free-insertion libration orbit around the second Lagrangian point of the Sun-Earth system (L2). Based on the associated quaternion computations, the key parameters described in section 4 are derived to compare S/C behavior with its limitations. The following figures give the statistics of the spacecraft key parameters in the case of the implemented reference survey.

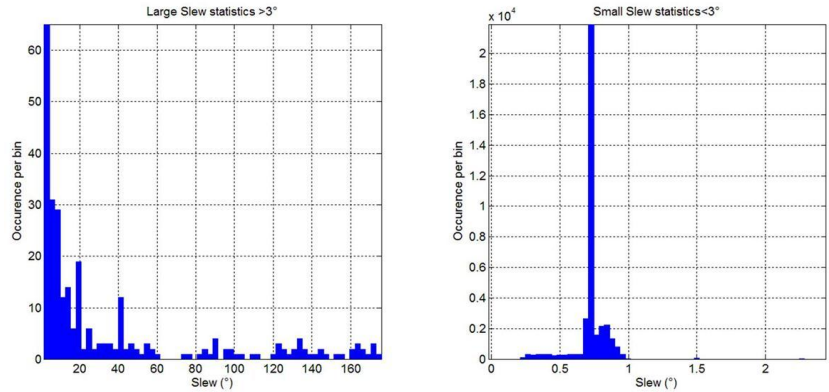

Figure 5. Statistics of large and small slews range.

The total number of fields of the survey (including calibration) is 41 500. The average size of a large slew is 40°, and there are 275 large slews operations. This is compatible with a maximum of 36x6 large slews of 180°.

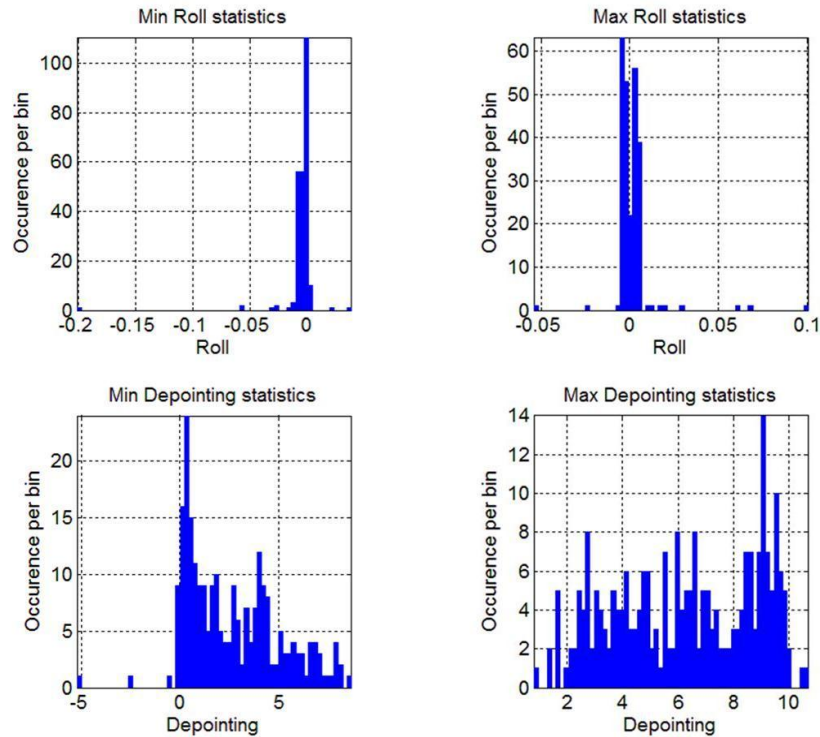

Figure 6. Min and Max Roll (shall be less than 5°) and Depoiniting (shall be less than 10°) angles distributions for the full survey.

The full survey is performed within Solar Aspect Angle variations allowed by the spacecraft for all fields (figure 7) and from one field to the next (figure 8). Not all the pointing capabilities of the spacecraft have been used for this survey implementation (limited roll), and therefore further optimization room is left for improving implementation efficiency.

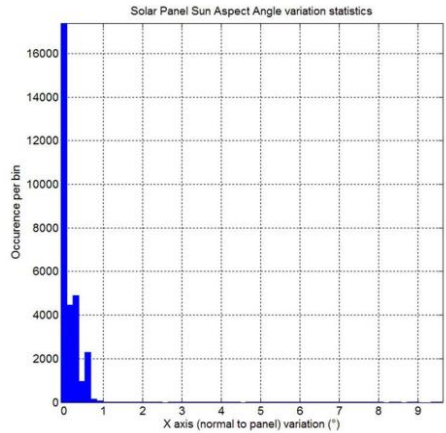

Figure 7. Solar panel variation from one field to the next along the full survey (shall be less than 10°)

These results show that the reference survey is compatible with the current design constraints from the spacecraft.

The duration of the survey is illustrated in figure 8, where the total area of the Wide survey is given by the blue line. The green line shows the part of the total survey which is performed over the core preferred area. The implemented strategy clearly observes primarily the best available square degrees during the first 4 years before extending the survey excursion toward the extended preferred area as to improve efficiency.

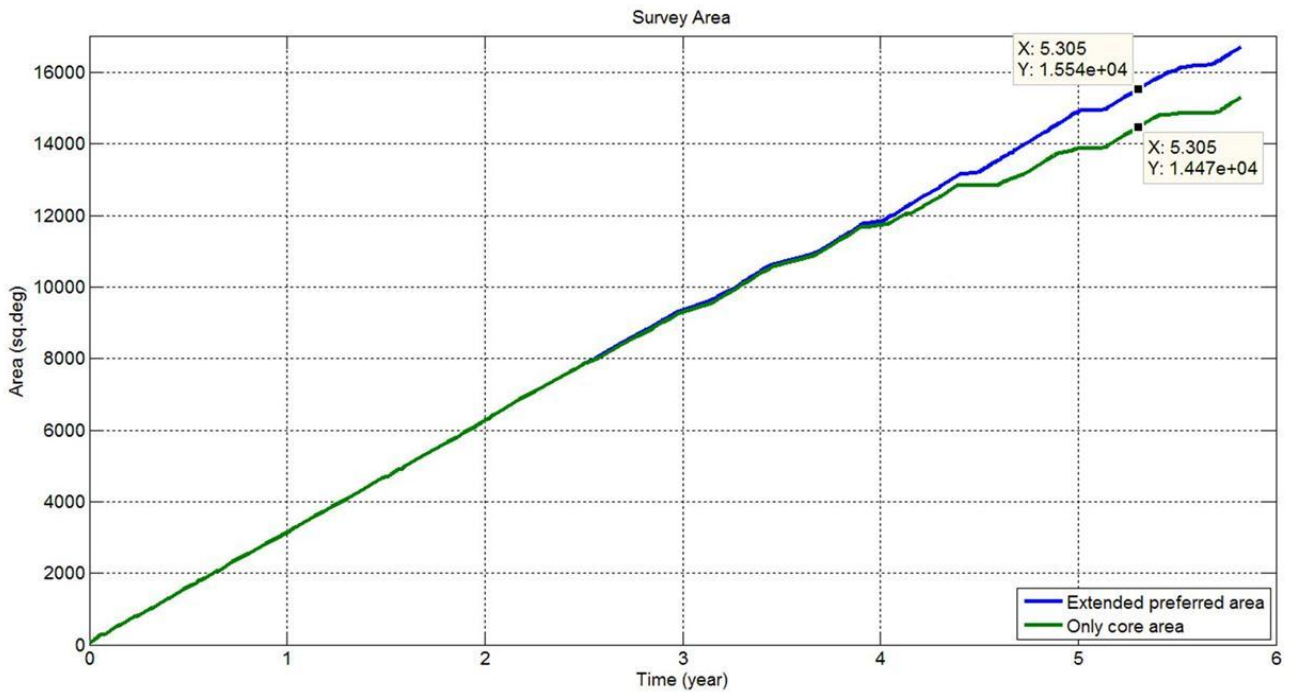

Figure 8. Survey area coverage as a function of survey time elapsed since beginning of nominal science operations. The green line is the area covered over the core preferred area, the blue line shows the total area survey, (core + extended preferred area).

The result shows a very high efficiency of the survey, and after 5.5 years (corresponding to 5.3 year on the plotted line + 2.5 months for orbital maneuvers) the total wide survey covered area is 15 500 deg², out of which 14 500 deg² are within the core preferred area. The 40 deg² of the Deep survey have been observed 2 magnitudes deeper than the wide and all required calibration data have been gathered.

# 6. FUTURE PROGRESS FOR THE *EUCLID* SURVEY

It has been a critical exercise to demonstrate that the *Euclid* survey can be implemented, taking into account the calibration requirements and the spacecraft constraints. The final results show that the 15 000 deg² coverage can be reached in 5.5 years at the beginning of the implementation phase. This leaves a 6 month margin with respect to the nominal lifetime of 6 years after the commissioning and science verification phases.

Now that the feasibility of this survey is demonstrated, the *Euclid* Sky Survey Working Group will focus on the optimization of the survey implementation. This will be done by building and examining in great details scenarios that are compatible with the same constraints, so to ensure optimal return to the community for both the primary and legacy science parts of the mission.